\documentclass[aps,english,showpacs,twocolumn,prb]{revtex4-1}
\usepackage{amsfonts}
\usepackage{amssymb}
\usepackage{amsmath}
\usepackage{graphicx}
\usepackage{epsfig}
\usepackage{color}
\usepackage{float}
\usepackage{subfigure}

\begin{document}

\title{Transition from degeneracy to coalescence: theorem and applications}
\author{P. Wang}
\author{K. L. Zhang}
\author{Z. Song}
\email{songtc@nankai.edu.cn}
\affiliation{School of Physics, Nankai University, Tianjin 300071, China}
\begin{abstract}
Exceptional point (EP) is exclusive for non-Hermitian system and distinct
from that at a degeneracy point (DP), supporting intriguing dynamics, which
can be utilized to probe quantum phase transition and prepare eigenstates in
a Hermitian many-body system. In this work, we investigate the transition
from DP for a Hermitian system to EP driven by non-Hermitian terms. We
present a theorem on the existence of transition between DP and EP for a
general system. The obtained EP is robust to the strength of non-Hermitian
terms. We illustrate the theorem by an exactly solvable quasi-one
dimensional model, which allows the existence of transition between fully
degeneracy and exceptional spectra driven by non-Hermitian tunnelings in
real and $k$ spaces, respectively. We also study the EP dynamics for
generating coalescing edge modes in Su-Schrieffer-Heeger-like models. This
finding reveals the ubiquitous connection between DP and EP.
\end{abstract}

\maketitle



\section{Introduction}

\label{Introduction}

The theoretical\cite{bender2007making,moiseyev2011non,krasnok2019anomalies}
and experimental studies\cite%
{guo2009observation,ruter2010observation,peng2014parity,feng2014single,hodaei2014parity,feng2017non,longhi2018parity,el2018non,miri2019,ozdemir2019parity,wu2019observation}
on the non-Hermitian system\ indicate that the interplay of lattice geometry
and non-Hermitian elements, such as imaginary on-site potential \cite%
{jin2013scaling,jin2011hermitian} and asymmetry tunneling\cite{XZZhangAP2013}
can induce exotic quantum dynamics \cite%
{PRL3,longhi2017unidirectional,XQLPRA2015,LJinPRL2018,LJinCPL2021}, which
never happens in a Hermitian system. Intuitively, these phenomena are known
to arise from the appearance of complex eigen energy within symmetry broken
region, leading to the explosion of the Dirac probability of the eigenstate.
However, it follows from a peculiar feature of the non-Hermitian system,
exceptional point (EP) dynamics, without the need of symmetry breaking. The
EP in a non-Hermitian system occurs when two or more eigenstates coalesce,
and usually associates with the non-Hermitian phase transition\cite{miri2019}%
. In a parity-time ($\mathcal{PT}$) symmetric non-Hermitian system (or other
similar systems), the $\mathcal{PT}$-symmetry of eigenstates spontaneously
breaks at the EP, which determines the exact $\mathcal{PT}$-symmetric phase
and the broken $\mathcal{PT}$-symmetric phase in this system. The EP plays a
pivotal role in intriguing dynamics and applications including asymmetric
mode switching\cite{Doppler2016}, unidirectional lasing \cite%
{Ramezani2014,Peng2016,LJinPRL2018}, and enhanced optical sensing\cite%
{Wiersig2014,Liu2016,Hodaei2017,Chen2017,Lau2018,Zhang2019,Lai2019,Hokmabadi2019}%
. Recently, EP dynamics is employed to engineer a target quantum state\cite%
{TEL2014,CL2015,XMY2020,ZXZ2020} and probe quantum phase transitions \cite%
{ZKLprl}. The mechanism of such a scheme is setting the target state as
coalescing state of a non-Hermitian Hamiltonian. In general, a familiar
target state as the central resource of quantum information processing is
always an eigenstate of Hermitian system, such as topological state,
many-particle entangled state. It, therefore, requires the coalescing state
is Hermitian-related and robust to the perturbation.

The aim of this paper is to provide a method for setting a target quantum
state to be a coalescing state. As well known, an EP is exclusive for
non-Hermitian system and is distinct from a degeneracy point (DP). We will
show that the transition from DP for a Hermitian system to EP can be
realized by proper non-Hermitian terms. This theorem indicates that, for a
general Hermitian system with a DP, one of degeneracy eigenstates can be set
as a coalescing state of a non-Hermitian system.\ This allows the scheme to
prepare a target quantum state based on the EP dynamics. To illustrate the
theorem, we investigate an exactly solvable quasi-one dimensional model,
which supports the transition between fully degeneracy and exceptional
spectra. It is shown that such a model in the thermodynamic limit is
equivalent to a two-coupled ring with power-law decay long-range hopping
term. As its application, we also study the EP dynamics for generating
coalescing edge modes in Su-Schrieffer-Heeger (SSH) like models. This
finding provides a way to construct a non-Hermitian system with EP based on
a Hermitian system with DP. The advantage of this method is that the
coalescing state is the eigenstate of a Hermitian system and is robust to
the strength of the non-Hermitian term. We expect our results benefit to
experimental research.

This paper is organized as follows. In Sec. \ref{Robust EP from DP}, we
present a theorem for a general system to create robust EP from DP. In Sec. %
\ref{Zero point in real space} and \ref{Zero point in k-space} we illustrate
the theorem from two examples. In Sec. \ref{Emerging edge modes}, we propose
a dynamic scheme to prepare edge modes as application of the theorem.
Finally, we provide a summary in Sec. \ref{Summary}.

\section{Theorem on robust EP from DP}

\label{Robust EP from DP}

A general system at EP is obtained by tuning an imaginary parameter, such as
imaginary potential or flux, to switch real energy levels to complex ones.
The critical value of the parameter is usually the solution of a
transcendental equation, and the EP system is sensitive to the imaginary
parameter. Then is a little tough to set an EP system precisely in practice.
The main aim of this paper is to answer the questions of whether a robust EP
system can be obtained. In this section, we present a theorem on
establishing EP based on DP of a Hermitian system. We will show that the
obtained EP is not sensitive to the strength of the non-Hermitian parameters.

We consider a general non-Hermitian Hamiltonian in the form
\begin{equation}
H=H_{0}+H^{\prime },
\end{equation}%
which can be separated into two parts, a Hermitian and non-Hermitian ones,
i.e.,
\begin{equation}
H_{0}=\left( H_{0}\right) ^{\dag },\text{but }H^{\prime }\neq \left(
H^{\prime }\right) ^{\dag }.
\end{equation}%
Here $H_{0}$ has two-fold degenerate eigenstates $\left\vert A\right\rangle $
and $\left\vert B\right\rangle $, i.e.,%
\begin{equation}
H_{0}\left\vert A\right\rangle =0,H_{0}\left\vert B\right\rangle =0,
\end{equation}%
where we take zero degenerate eigen energy for the sake of simplicity.\ If
the non-Hermitian term $H^{\prime }$ satisfies%
\begin{equation}
H^{\prime }\left\vert A\right\rangle =0,\left( H^{\prime }\right) ^{\dag
}\left\vert B\right\rangle =0,
\end{equation}%
we have%
\begin{equation}
H\left\vert A\right\rangle =0,H^{\dag }\left\vert B\right\rangle =0.
\end{equation}%
Two states $\left\vert A\right\rangle $ and $\left\vert B\right\rangle $ are
mutually biorthogonal conjugate and $\langle B\left\vert A\right\rangle $\
is the biorthogonal norm of them. The vanishing norm indicates that state $%
\left\vert A\right\rangle $ is coalescing state of $H$.\ So when two
Hamiltonians $H_{0}$\ and $H^{\prime }$ ($\left( H^{\prime }\right) ^{\dag }$%
)\ have a common zero-energy state $\left\vert A\right\rangle $ ($\left\vert
B\right\rangle $), one can say that Hamiltonian $H$\ get an EP, which is
robust to the strength of $H^{\prime }$\ varies.

The theorem is given here without specific reference to the detailed form of
the Hamiltonian. It should work for the nonrelativistic and relativistic,
continuous and discrete Hamiltonians. Applying it to a tight-binding model,
we can find some detailed signatures of the degenerate eigen states.
Considering the conditions, $H^{\prime }\left\vert A\right\rangle =0$ and $%
\left( H^{\prime }\right) ^{\dag }\left\vert B\right\rangle =0$, the
simplest example for such a $H^{\prime }$ is unidirectional hopping term,
i.e., $\kappa a_{i}^{\dag }a_{j}$ \ ($i\neq j$), where $a_{i}$ and $a_{j}$\
are fermion or boson operators. Note that the above conditions can be
satisfied if\ $a_{j}\left\vert A\right\rangle =0$ and $a_{i}\left\vert
B\right\rangle =0$. It means that two states $\left\vert A\right\rangle $
and $\left\vert B\right\rangle $\ have nodal points at $j$ and $i$
respectively. In addition, the existence of EP is independent of the nonzero
value of $\kappa $. This rigorous conclusion has important implications in
the design of quantum devices to prepare a target quantum state at will. In
the following sections, we will present serval illustrative examples to
demonstrate the theorem.

\section{Zero point in real space}

\label{Zero point in real space}

\begin{figure*}[t]
\centering
\includegraphics[bb=0 0 1560 524, width=17 cm, clip]{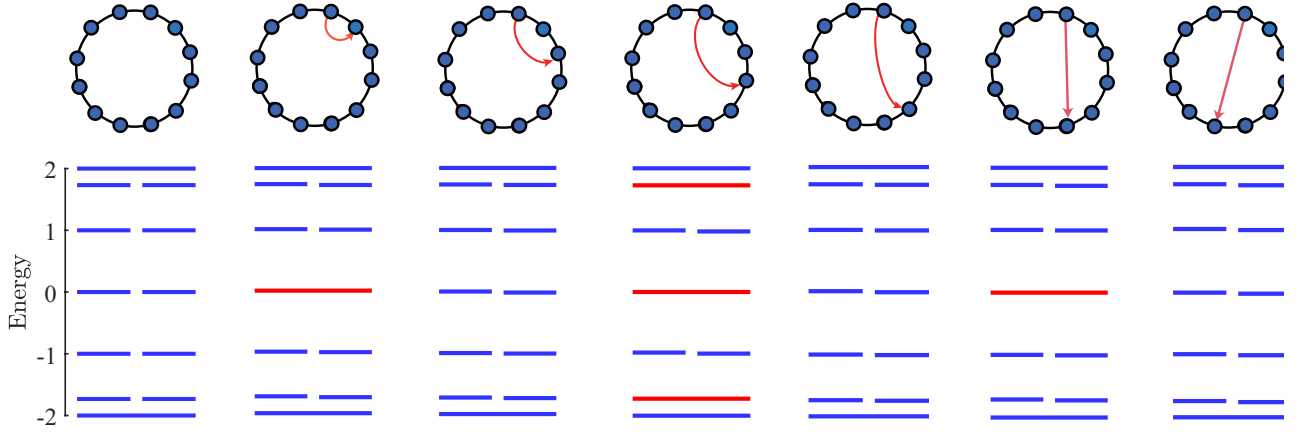}
\caption{{Schematic illustration of the levels of a Hermitian $12$-site
ring, which show the effect of an additional single unidirectional hopping
(red arrow) crossing two sites. Two degenerate levels are indicated by a
pair of blue segments, while a coalescing level is in red. The parameter for
red arrow is $\protect\kappa =0.5$. We can see that a single asymmetric
hopping term for a Hermitian ring can drive the transition between DP and EP
for some specific but not any levels. }}
\label{fig1}
\end{figure*}
Consider a uniform $2N$-site ring system with the Hamiltonian%
\begin{equation}
H_{0}=\sum_{j=1}^{2N}(c_{j}^{\dag }c_{j+1}+\text{\textrm{H.c.}}%
)=2\sum_{k}\cos kc_{k}^{\dag }c_{k},
\end{equation}%
where%
\begin{equation}
c_{k}^{\dag }=\frac{1}{\sqrt{2N}}\sum_{j=1}^{2N}e^{ikj}c_{j}^{\dag },
\end{equation}%
with the wave vector $k=\pi n/N$, $n=1,2,...,2N$. We note that there are $%
N-1 $ pairs of degenerate eigenstates. The aim of this section is to answer
the question of what kind of $H^{\prime }$ can result in the transition from
a DP to EP.\ A single-particle eigenstate has the form%
\begin{equation}
\left\vert k\right\rangle =\frac{1}{\sqrt{2N}}\sum_{j=1}^{2N}e^{ikj}\left%
\vert j\right\rangle ,
\end{equation}%
which has no nodal point for any $k$. Here $\left\vert j\right\rangle
=c_{j}^{\dag }\left\vert 0\right\rangle $ is the position state, where $%
\left\vert 0\right\rangle $\ is the vacuum state. However, we can construct
two-fold degenerate eigenstates in the form%
\begin{equation}
\left\vert \psi _{k}^{\pm }\right\rangle =\frac{1}{\sqrt{2}}\left(
\left\vert k\right\rangle \pm e^{i2kl_{0}}\left\vert -k\right\rangle \right)
,
\end{equation}%
with $\langle \psi _{k}^{+}\left\vert \psi _{k}^{-}\right\rangle =0$. The
additional factor $e^{i2kl_{0}}$\ leads to a nodal point at $l_{0}$th site,
i.e.,%
\begin{equation}
\langle l_{0}\left\vert \psi _{k}^{-}\right\rangle =0.
\end{equation}%
It easy to check that for another state $\left\vert \psi
_{k}^{+}\right\rangle $, there is a nodal point at the $\left(
l_{0}+r\right) $th site, i.e.,
\begin{figure*}[t]
\centering
\includegraphics[bb=0 0 1132 454, width=0.80\textwidth]{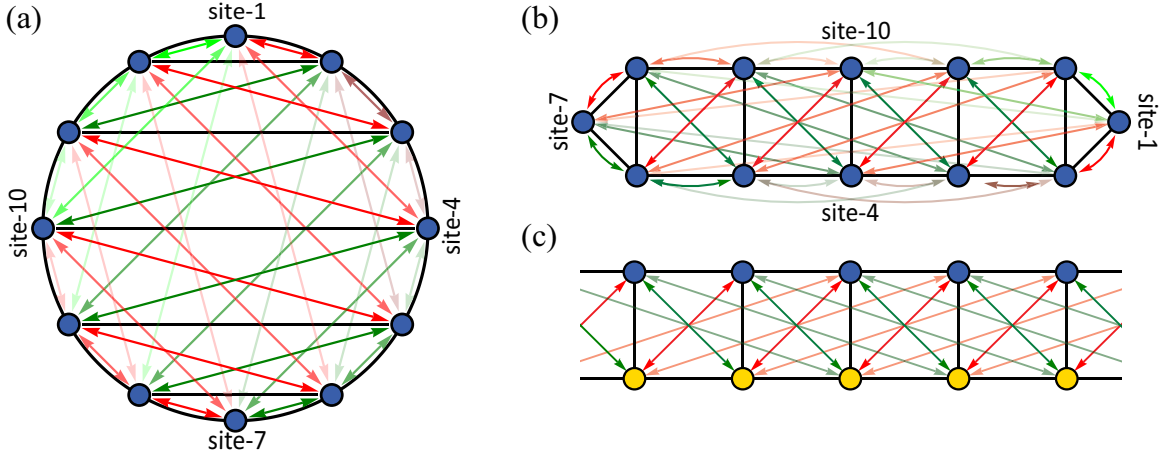}
\caption{(a) The schematic diagram of Hamiltonian in {Eq. }(\protect\ref{H2'}%
). The black straight lines denote the real long-range couplings. The
colored lines with bidirectional arrows denote the pure imaginary couplings,
and we adopt different colors and opacity to distinguish the imaginary
couplings. The dark red lines represent the coupling where the sum of the
site's position is equal to $5$ or $3$, and we further distinguish $5$ and $%
3 $ by the reduced transparency. The same goes for the green lines standing
for $23$, $21$, $19,$ the dark green lines for $11$, $9$, $7$, and the red
lines for $17$, $15$, $13$. (b) We reshape the schematic diagram of (a) to
be a ladder system. {(c) The schematic illustration of Hamiltonian in Eq. (%
\protect\ref{H3}). The black ladder represents }$H_{0}$, the colored lines
with bidirectional arrows denote $H^{\prime }$. Ignoring the hoppings on the
boundary, the intralayer long-range hoppings for (b) and (c) have the same
linking method but different strengths.}
\label{fig2}
\end{figure*}
\begin{equation}
\langle l_{0}+r\left\vert \psi _{k}^{+}\right\rangle =0,
\end{equation}%
when the following condition is satisfied%
\begin{equation}
\cos \left( kr\right) =0,
\end{equation}%
which requires $k$\ to be specific values%
\begin{equation}
k=\frac{\left( 2m+1\right) \pi }{2r},m=0,1,2,....
\end{equation}%
Then non-Hermitian term $H^{\prime }$ is in the form%
\begin{equation}
H^{\prime }=\kappa c_{l_{0}}^{\dag }c_{l_{0}+r}.
\end{equation}%
$\allowbreak \allowbreak $ $\allowbreak \allowbreak $For instance, for $r=1$%
, we have $k=\pi /2$, and for $r=2$, we have $k=\pi /4$ and $k=3\pi /4$. It
means that one can take $H^{\prime }=\kappa c_{l_{0}}^{\dag }c_{l_{0}+1}$ to
acquire the coalescing state $\left\vert \psi _{\pi /2}^{+}\right\rangle $,
while taking $H^{\prime }=\kappa c_{l_{0}}^{\dag }c_{l_{0}+2}$, obtain $%
\left\vert \psi _{3\pi /4}^{+}\right\rangle $. In Fig. \ref{fig1}, we
schematically illustrate a finite system with several kinds of non-Hermitian
hopping terms and plot the corresponding energy levels. It indicates that
for the finite system, there are some DPs that cannot be transmitted to EPs
by a single asymmetric hopping term.

\section{Zero point in $k$-space}

\label{Zero point in k-space}

In the previous section, we have shown that a single asymmetric hopping term
for a Hermitian ring can drive the transition between DP and EP for some
specific but not any levels. A natural question is what kind of $H^{\prime }$
can result in the transition from a DP to EP for any given $k$. Actually,
this can be done by simply taking $H_{k}^{\prime }=Jc_{k}^{\dag }c_{-k}$,
which contains all range unidirectional hopping terms. In the above example,
$H^{\prime }=\kappa c_{i}^{\dag }c_{j}$, we do not restrict $c_{i}^{\dag }$
and $c_{j}$\ to be the operators in real space. For example, a fermionic EP
Hamiltonian can be $H^{\prime }=\kappa c_{k_{1}}^{\dag }c_{k_{2}}$. The
corresponding coalescing state is $c_{k_{1}}^{\dag }\left\vert
0\right\rangle _{k_{1}}\left\vert 0\right\rangle _{k_{2}}$, while the
auxiliary state is $c_{k_{2}}^{\dag }\left\vert 0\right\rangle
_{k_{1}}\left\vert 0\right\rangle _{k_{2}}$.

Now we investigate an example to demonstrate the application of the result
above. We consider a system with the non-Hermitian term%
\begin{equation}
H^{\prime }=\sum_{k}H_{k}^{\prime }=\sum_{k>0}c_{k}^{\dag }c_{-k}.
\end{equation}%
In the case of $k\neq 0$\ and $\pi $, both $H^{\prime }\left\vert
k\right\rangle =0$\ and $\left( H^{\prime }\right) ^{\dagger }\left\vert
-k\right\rangle =0$\ always hold, which indicates $H^{\prime }$\ drives all
DPs into EPs. Straightforward derivation leads to%
\begin{eqnarray}
H_{k}^{\prime } &\approx &\frac{1}{N}\underset{l+j=odd(l>j)}{\sum }i\cot [%
\frac{(l+j)\pi }{N}]\left( c_{l}^{\dagger }c_{j}+\mathrm{H.c.}\right)
\notag \\
&&+\frac{1}{2}\underset{l+j=N,2N}{\sum }c_{l}^{\dag }c_{j}.  \label{H2'}
\end{eqnarray}%
In Fig. \ref{fig2}(a), we schematically illustrate a finite system with
above $H^{\prime }$. We note that the non-Hermitian hopping strength $i\cot
[(l+j)\pi /N]/N$ is large as $l+j$ close to $0$, $N$, and $2N$. For $l+j=N+1$%
, we have $l-j=N+1-2j$, which indicates long range hopping. For example,
taking $j=N/4$, we have $l=3N/4+1$. Then the hopping spacing is $l-j=N/2+1$.
However, if we reshape the ring as a ladder, such long-range hopping terms
become short-range as shown in Fig. \ref{fig2}(b). In large $N$\ limit,
considering the case with $l+j=nN+\Delta $, with $\Delta \ll N$\ $(\Delta
=\pm 1,\pm 3,...,$\ and $n=0,1,2,...),$\ we can simplify the coupling
constant%
\begin{equation}
\frac{1}{N}\cot [\frac{(l+j)\pi }{N}]=\frac{1}{N}\cot [\frac{\Delta \pi }{N}%
]\approx \frac{1}{\pi \Delta }.  \label{appr}
\end{equation}%
This example indicates that for the finite system, $N-1$ pairs of degenerate
levels can be switched into coalescing $N-1$ levels by $H^{\prime }$.

The underlying mechanism of realizing the transition from degeneracy
spectrum to coalescing spectrum can be revealed by the following model.
\textbf{Inspired by Eq. (\ref{appr}), }we consider a two coupled uniform
chains system, which has a two-leg ladder structure. Figure \ref{fig2}(c)
sketches the geometry of the system, in which the hopping amplitudes in each
leg are uniform and the hopping strengths are distance-dependent. Such a
ladder system is a bipartite lattice system, consisting of two sub-lattices $%
A$ and $B$. We write down the Hamiltonian for the system in the form%
\begin{eqnarray}
H_{0} &=&\sum_{j=1}^{N}(a_{j}^{\dagger }b_{j}+a_{j}^{\dagger
}a_{j+1}+b_{j}^{\dagger }b_{j+1}+\mathrm{H.c.})\mathrm{,}  \label{H3} \\
H^{\prime } &=&\frac{iJ}{2}\sum_{j=1}^{N}\sum_{n=1}\frac{1}{2n-1}%
(a_{j}^{\dagger }b_{j+2n-1}-b_{j}^{\dagger }a_{j+2n-1}+\mathrm{H.c.})\mathrm{%
,}  \notag
\end{eqnarray}%
where $a_{l}^{\dag }$ and $b_{l}^{\dag }$ are the creation operators of
fermion or boson at the $l$th site of sub-lattice $A$ and $B$, respectively.
We take a periodic boundary condition, by setting $a_{N+1}^{\dagger
}=a_{1}^{\dagger }$ and $b_{N+1}^{\dagger }=b_{1}^{\dagger }$. Taking the
transformation%
\begin{equation}
\left\{
\begin{array}{c}
a_{k}=\frac{1}{\sqrt{N}}\sum_{j}e^{ikj}a_{j} \\
b_{k}=\frac{1}{\sqrt{N}}\sum_{j}e^{ikj}b_{j}%
\end{array}%
\right. ,
\end{equation}%
we have
\begin{equation}
H=\sum_{k}H_{k}=\sum_{k}(a_{k}^{\dagger },b_{k}^{\dagger })h_{k}\left(
\begin{array}{c}
a_{k} \\
b_{k}%
\end{array}%
\right) ,
\end{equation}%
where the wave vector $k=\pi (2n-N)/N$, $(n=0,1,...,N-1)$. The Bloch
Hamiltonian is%
\begin{equation}
h_{k}=\left(
\begin{array}{cc}
0 & 1-\Delta _{k} \\
1+\Delta _{k} & 0%
\end{array}%
\right) +2\cos k.
\end{equation}%
where
\begin{equation}
\Delta _{k}=J\sum_{n=1}\frac{\sin \left[ \left( 2n-1\right) k\right] }{2n-1}.
\end{equation}%
The Hamiltonian $H$ can be easily diagonalized since $[H_{k},H_{k^{\prime
}}]=0$, and the spectrum is%
\begin{equation}
\varepsilon _{k}=2\cos k\pm \sqrt{1-\left( \Delta _{k}\right) ^{2}}.
\end{equation}%
We note that when the summation in $\Delta _{k}$\ covers to infinity, $%
\Delta _{k}$ is a step function%
\begin{equation}
\Delta _{k}=\frac{J\pi }{4}\mathrm{sgn}(k)
\end{equation}%
according to Fourier analysis. Then the spectrum becomes%
\begin{equation}
\varepsilon _{k}=\pm \sqrt{1-\left( J\pi /4\right) ^{2}},
\end{equation}%
which vanishes at $J=4/\pi $ and the EP spectrum appears.

\section{Emerging edge modes}

\label{Emerging edge modes}

In the above sections, the obtained coalescing states are all extended
states, which have real wave vectors. In this section, we focus on the
transition from DP to EP where the coalescing bound states appear. A
coalescing bound state is a local state and lives at an energy gap, and then
can be a stable target state of the time evolution at EP. In the following,
we at first present two examples of coalescing edge states in SSH-like
models. Then study the dynamical preparation of a $2$D edge state.

\subsection{SSH chain}

\label{SSH chain}

\begin{figure}[t]
\centering
\includegraphics[bb=0 0 628 675, width=8 cm, clip]{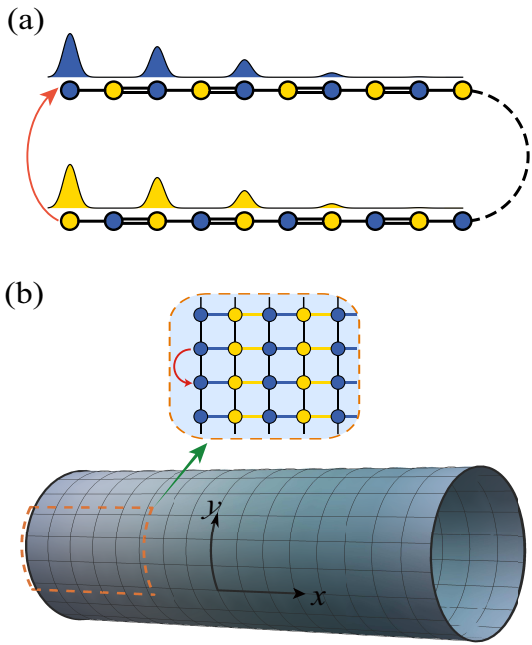}
\caption{{}(a) Schematic illustration of SSH chain. The unidirectional {%
hopping} $\protect\kappa $ pointing from the head to the tail of the SSH
chain is denoted by a red arrow. Two degenerate edge states $\left\vert
R\right\rangle $ (yellow) and $\left\vert L\right\rangle $ (blue) appear
when taking $\protect\kappa =0$. When $\protect\kappa \neq 0$, there only
exists a coalescing state $\left\vert L\right\rangle $. (b) Schematic
illustration of SSH cylinder. The details of couplings are shown in the
inset, wherein the unidirectional coupling is indicated by the red arrow.}
\label{fig3}
\end{figure}

We start our investigation by considering a SSH chain with single
unidirectional hopping across two ends, with the Hamiltonian in the form%
\begin{eqnarray}
H_{0} &=&\frac{1}{2}\sum_{l=1}^{2N-1}\left[ 1+\left( -1\right) ^{l}\delta %
\right] c_{l}^{\dagger }c_{l+1}+\text{\textrm{H.c.}},  \notag \\
H^{\prime } &=&\kappa c_{1}^{\dag }c_{2N}.
\end{eqnarray}%
It is a bipartite lattice, i.e., it has two sublattices $A$, $B$ such that
each site on lattice $A$ has its nearest neighbors on sublattice $B$, and
vice versa.\textbf{\ }The Hermitian system $H_{0}$ is the prototype of a
topologically nontrivial band insulator with a symmetry protected
topological phase \cite{Ryu,Wen}. In recent years, it has been attracted
much attention and extensive studies have been demonstrated \cite%
{XD,Hasan,Delplace,ChenS1,ChenS2,LS PRA}. The degenerate zero modes take the
role of topological invariant in the infinite $N$ limit and are explicitly
expressed as%
\begin{equation}
\left\{
\begin{array}{c}
\left\vert \text{L}\right\rangle =\frac{1}{\sqrt{\Omega }}%
\sum\limits_{j=1}^{N}\left( \frac{\delta -1}{\delta +1}\right)
^{j-1}c_{2j-1}^{\dagger }\left\vert 0\right\rangle \\
\left\vert \text{R}\right\rangle =\frac{1}{\sqrt{\Omega }}%
\sum\limits_{j=1}^{N}\left( \frac{\delta -1}{\delta +1}\right)
^{N-j}c_{2j}^{\dagger }\left\vert 0\right\rangle%
\end{array}%
\right. ,
\end{equation}%
for $\delta >0$\ where the normalization factor is $\Omega =\{1-\left[
\left( \delta -1\right) /\left( \delta +1\right) \right] ^{2N}\}/\{1-\left[
\left( \delta -1\right) /\left( \delta +1\right) \right] ^{2}\}$. It is easy
to check that%
\begin{equation}
c_{2N}\left\vert \text{L}\right\rangle =c_{1}\left\vert \text{R}%
\right\rangle =0,
\end{equation}%
which ensues that%
\begin{equation}
H^{\prime }\left\vert \text{L}\right\rangle =\left( H^{\prime }\right)
^{\dag }\left\vert \text{R}\right\rangle =0.
\end{equation}%
According to the theorem aforementioned, $\left\vert \text{L}\right\rangle $
is a coalescing edge state. In previous work \cite{CL2016}, it has been
shown that, two bound states appear in the bulk of a chain when a strong
bond is replaced by a \textbf{tiny} one. Now we take it as a unidirectional
hopping as described in $H^{\prime }$. In this sense\textbf{, }it is not
surprising that a single bound state $\left\vert \text{L}\right\rangle $ can
appear in the bulk of a finite size but sufficient long chain. The profiles
of the edge modes are schematically illustrated in Fig. \ref{fig3} (a). In
contrast, when the unidirectional hopping replaces a weak bond, there is no
zero mode appears. Therefore, the topological feature can be demonstrated by
adding a non-Hermitian impurity, as a generalization of the bulk-edge
correspondence.

\subsection{SSH cylinder}

\label{SSH cylinder}

A similar situation can happen in 2D system by adding a local non-Hermitian
impurity. In the following we present an example of 2D system, which is an
extended 2D SSH cylinder. The 2D SSH cylinder consists of $M$ (even) chains
which are uniformly coupled,%
\begin{eqnarray}
H_{\mathrm{sc}}^{0} &=&\sum_{j=1}^{M}\sum_{l=1}^{2N-1}\left[ 1+\left(
-1\right) ^{l}\delta \right] c_{j,l}^{\dagger }c_{j,l+1}  \notag \\
&&+J\sum_{j=1}^{M}\sum_{l=1}^{2N}c_{j,l}^{\dagger }c_{j+1,l}+\text{\textrm{%
H.c.}},
\end{eqnarray}%
and a perturbation on the boundary,%
\begin{equation}
H_{\mathrm{sc}}^{\prime }=\kappa c_{1,1}^{\dagger }c_{M,1},
\label{CylinderNonHer}
\end{equation}%
where $j$ ($l$) is the index of row (column). Therefore, the Hamiltonian of
the SSH cylinder reads $H_{\mathrm{sc}}=H_{\mathrm{sc}}^{0}+H_{\mathrm{sc}%
}^{\prime }$. Figure \ref{fig3}(b) shows the schematic illustration of SSH
cylinder, the red arrow in the inset represents the perturbation. $H_{0}^{%
\mathrm{sc}}$ has four degenerate local zero modes, and two of which
localizes at the left boundary
\begin{equation}
\left\{
\begin{array}{c}
\left\vert L_{\mathrm{e}}\right\rangle =\frac{1}{\sqrt{\Omega }}%
\sum\limits_{j=1}^{M/2}\sum\limits_{l=1}^{N}\left( \frac{\delta -1}{\delta +1%
}\right) ^{l-1}c_{2j,2l-1}^{\dagger }\left\vert 0\right\rangle \\
\left\vert L_{\mathrm{o}}\right\rangle =\frac{1}{\sqrt{\Omega }}%
\sum\limits_{j=1}^{M/2}\sum\limits_{l=1}^{N}\left( \frac{\delta -1}{\delta +1%
}\right) ^{l-1}c_{2j-1,2l-1}^{\dagger }\left\vert 0\right\rangle%
\end{array}%
\right. ,
\end{equation}%
where $\Omega =M/2\{1-\left[ \left( \delta -1\right) /\left( \delta
+1\right) \right] ^{N}\}/\{1-\left[ \left( \delta -1\right) /\left( \delta
+1\right) \right] ^{2}\}$. It is not hard to check that%
\begin{equation}
H_{\mathrm{sc}}^{\prime }\left\vert L_{\mathrm{o}}\right\rangle =0,\left( H_{%
\mathrm{sc}}^{\prime }\right) ^{\dagger }\left\vert L_{\mathrm{e}%
}\right\rangle =0,
\end{equation}%
which means $\left\vert L_{\mathrm{o}}\right\rangle $\ is a coalescing edge
state of $H_{\mathrm{sc}}$, and $\left\vert R_{\mathrm{e(o)}}\right\rangle $%
\ remains degenerate states, $H\left\vert R_{\mathrm{e(o)}}\right\rangle =0.$

\subsection{Dynamical preparation of edge state}

\label{Dynamical preparation of edge state}

It seems that DP and EP systems are totally two different ones. Taking $%
H^{\prime }\rightarrow \kappa H^{\prime }$ to impose a strength on the
non-Hermitian term, one can investigate the effect of $H^{\prime }$ on the
system quantitatively. Intuitively, a small change of $\kappa $ from zero
can result in a drastic change. However, in the following, we will show that
there exists a continuous crossover between them. We measure the signature
of the system by detecting the dynamics of the observable, such as the time
evolution of particle probability. As the application of the theorem, the
relationship between time and particle probability reminds a new approach to
prepare edge modes.

We consider a $2$-site system as the simplest example for the obtained
theorem, which has the Hamiltonian%
\begin{equation}
H_{2\text{s}}=\kappa c_{1}^{\dag }c_{2}+\varepsilon _{0}\left( c_{1}^{\dag
}c_{1}+c_{2}^{\dag }c_{2}\right) .
\end{equation}%
The time evolution operator has the form
\begin{eqnarray}
U(t) &=&\exp (-iH_{2\text{s}}t)  \notag \\
&=&\exp \left[ -i\varepsilon _{0}\left( c_{1}^{\dag }c_{1}+c_{2}^{\dag
}c_{2}\right) t\right] \exp (-i\kappa c_{1}^{\dag }c_{2}t),
\end{eqnarray}%
which drives the time evolution for an initial state $\left\vert \psi
(0)\right\rangle $%
\begin{equation}
\left\vert \psi (t)\right\rangle =U(t)\left\vert \psi (0)\right\rangle .
\end{equation}

\begin{figure}[t]
\centering
\includegraphics[bb=0 0 491 701, width=8 cm, clip]{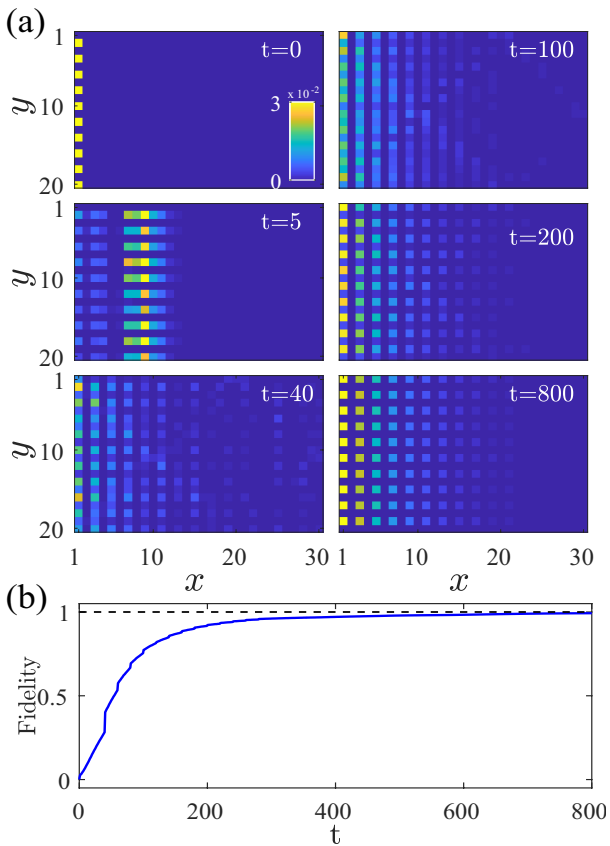}
\caption{(a) Snapshots of the probabaility distribution at various time
moments for initial excitation in Eq. {(\protect\ref{InitialState}).} The
probabaility is normalized at any moments. The system size is $M=20$, $N=100$%
. Only the limited region within $N=30$ is shown because the intensity is
almost vanishing outside this area. The other parameters are $\protect\delta %
=0.1,J=1,$ and $\protect\kappa =0.5$. (b) The fidelity between the evolved
state $\left\vert \protect\psi (t)\right\rangle $ and the coalescing edge
state $\left\vert L_{\mathrm{o}}\right\rangle $. The units of time is $1/J$.}
\label{fig4}
\end{figure}

(i) In the case of $\kappa =0$, we have%
\begin{equation}
\left\vert \psi (t)\right\rangle =\exp \left[ -i\varepsilon _{0}\left(
c_{1}^{\dag }c_{1}+c_{2}^{\dag }c_{2}\right) t\right] \left\vert \psi
(0)\right\rangle .
\end{equation}%
Then for an initial state with fixed particle number, i.e, $\left(
c_{1}^{\dag }c_{1}+c_{2}^{\dag }c_{2}\right) \left\vert \psi
(0)\right\rangle =n\left\vert \psi (0)\right\rangle $ with $n=0$, $1$, and $%
2 $, $U(t)$ only contributes a phase factor to $\left\vert \psi
(0)\right\rangle $.

(ii) In the case of $\kappa \neq 0$, we have%
\begin{equation}
\exp (-i\kappa c_{1}^{\dag }c_{2}t)=1-i\kappa c_{1}^{\dag }c_{2}t.
\end{equation}%
Then the time evolution of the coalescing state $\left\vert \psi _{\mathrm{c}%
}\right\rangle =c_{1}^{\dag }\left\vert 0\right\rangle $ is%
\begin{equation}
U(t)\left\vert \psi _{\mathrm{c}}\right\rangle =\exp \left( -i\varepsilon
_{0}t\right) \left\vert \psi _{\mathrm{c}}\right\rangle ,
\end{equation}%
while the time evolution of the auxiliary state $\left\vert \psi _{\mathrm{a}%
}\right\rangle =c_{2}^{\dag }\left\vert 0\right\rangle $ is%
\begin{equation}
U(t)\left\vert \psi _{\mathrm{a}}\right\rangle =\exp \left( -i\varepsilon
_{0}t\right) \left( \left\vert \psi _{\mathrm{a}}\right\rangle -i\kappa
t\left\vert \psi _{\mathrm{c}}\right\rangle \right) ,  \label{Hc}
\end{equation}%
where we have used the identity $\left( c_{1}^{\dag }c_{2}\right) ^{2}=0$.
The difference between DP and EP is obvious when $\kappa t\gg 1$. However,
within the time scale $t\ll 1/\kappa $, the dynamics under the DP and EP has
no difference. It indicates that the crossover from DP to EP is continuous.
Similarly, it has been shown that a non-Hermitian system around EP exhibits
some peculiar critical dynamics as EP \cite{XMY2020}.

Equation (\ref{Hc}) indicates that the time evolution of the coalescing
state linearly depends on time $t$, so the evolved state is approximately
equal to $\left\vert \psi _{\mathrm{c}}\right\rangle $ for the relatively
large time scale. When $\left\vert \psi _{\mathrm{c}}\right\rangle $ is a
localized coalescing state, the dynamical preparation of the robust edge
state is as follows. Consider the time evolution driven by Hamiltonian $H_{%
\mathrm{sc}}$, the initial state is%
\begin{equation}
\left\vert \psi (0)\right\rangle =\frac{1}{\sqrt{M/2}}\sum%
\limits_{j=1}^{M/2}c_{2j-1,1}^{\dagger }\left\vert 0\right\rangle ,
\label{InitialState}
\end{equation}%
which is just $\left\vert L_{\mathrm{e}}\right\rangle $ when $\delta $
infinitely approaches $1$. Figure \ref{fig4}(a) exhibits numerical
simulations of $\left\vert \psi (t)\right\rangle $, the system size is $%
N=100,M=20$. We only show the region $N\leqslant 30$ because the probability
is almost zero in the other region. Figure \ref{fig4}(b) exhibits the
fidelity%
\begin{equation}
F(t)=\left\langle L_{\mathrm{o}}\right. \left\vert \psi (t)\right\rangle
\end{equation}%
to evaluate the closeness of evolved state $\left\vert \psi (t)\right\rangle
$ and the coalescing edge state $\left\vert L_{\mathrm{o}}\right\rangle $.
In Fig. \ref{fig4}(a), we show the probability distribution of $\left\vert
\psi (0)\right\rangle $ in the real space, the fidelity is almost zero.
Before $t=100$, the probability distribution of $\left\vert \psi
(t)\right\rangle $ presents as an unstable stripe, and the fidelity has a
small value. At $t=200$, the stripe tends to be stable and looks like the
edge state, the fidelity reaches to around $0.9$. At $t=800$, we get a
stable stripe, and the fidelity is close to $1$ which indicates that $%
\left\vert \psi (t)\right\rangle $ is the coalescing edge state. Our
numerical results show that EP dynamics based scheme has better efficiency.
The setting of system parameters is expected to provide guidance for the
experiment.

\section{Summary}

\label{Summary}

In summary, we have developed a theory for a class of non-Hermitian
Hamiltonian which supports robust EPs. Such Hamiltonian consists two
separated parts, Hermitian and non-Hermitian ones. The Hermitian Hamiltonian
has degenerate eigenstates, which coalesce into a single state by the
non-Hermitian part. The most fascinating and important feature of such
systems is that the EP is not sensitive to the strength of the non-Hermitian
perturbation. As examples, we have investigate three types of systems: (i)
uniform ring system with a single asymmetric hopping term, in which several
pairs of degenerate states become coalescing states, (ii) uniform ladder
system with long-range power-law decaying imaginary hopping terms, in which
the degenerate spectrum becomes coalescing spectrum, (iii) SSH-like system
in a nontrivial topological phase with a single asymmetric hopping term, in
which the degenerate edge state becomes coalescing edge state. We also
demonstrate the application of the EP dynamics based on numerical
simulation. It is shown that the $2$D coalescing edge state can be generated
by a local initial state. Our findings offer a method for the efficient
construction of a robust EP system and are expected to be necessary and
insightful for quantum engineering.

\acknowledgments This work was supported by National Natural Science
Foundation of China (under Grant No. 11874225).

\end{document}